\documentclass[preprint,aps,floatfix]{revtex4}
\usepackage{amsmath}
\usepackage{graphicx}
\usepackage{epsfig}
\newlength{\upit}\upit=0.1truein

\newcommand{\ltappr}{{{\lower4pt\hbox{$<$} } \atop \widetilde{ \ \ \ }}}
\newlength{\bxwidth}\bxwidth=1.5 truein

\begin{document}
\newcommand{\dg}{^{\dagger }}
\newcommand{\vk}{\vec k}
\newcommand{\vq}{{\vec{q}}}
\newcommand{\vp}{\bf{p}}
\newcommand{\al}{\alpha}
\newcommand{\be}{\beta}
\newcommand{\si}{\sigma}
\newcommand{\rarrow}{\rightarrow}
\def\fig#1#2{\includegraphics[height=#1]{#2}}
\def\figx#1#2{\includegraphics[width=#1]{#2}}
\newlength{\figwidth}
\figwidth=10cm
\newlength{\shift}
\shift=-0.2cm
\newcommand{\fg}[3]
{
\begin{figure}[ht]

\vspace*{-0cm}
\[
\includegraphics[width=\figwidth]{#1}
\]
\vskip -0.2cm
\caption{\label{#2}
\small#3
}
\end{figure}}
\newcommand{\fgb}[3]
{
\begin{figure}[b]
\vskip 0.0cm
\begin{equation}\label{}
\includegraphics[width=\figwidth]{#1}
\end{equation}
\vskip -0.2cm
\caption{\label{#2}
\small#3
}
\end{figure}}

\newcommand \bea {\begin{eqnarray} }
\newcommand \eea {\end{eqnarray}}
\newcommand{\bk}{{\bf{k}}}
\newcommand{\bx}{{\bf{x}}}
{\large\bf{Quantum criticality}}\\
[0.5cm]

Piers Coleman$^{1}$ \& Andrew J. Schofield$^{2}$
\\[0.5cm]

{\em $^{1}$Center for Materials Theory, Rutgers University, Piscataway, New Jersey
08854-8019, USA}

{\em $^{2}$
School of Physics \&
Astronomy, University of Birmingham, Edgbaston, Birmingham B15 2TT, UK}\\[0.5cm]

{\sl \bf As we mark the centenary of Albert Einstein's seminal
contribution to both quantum mechanics and special relativity, we
approach another anniversary---that of Einstein's foundation of the
quantum theory of solids. But 100 years on, the same experimental
measurement that puzzled Einstein and his contemporaries is forcing us
to question our understanding of how quantum matter transforms at
ultra-low temperatures.}
\\

At the turn of the twentieth
century a crisis less grandiose, yet just as profound as Michelson and
Morley's failure to determine the Earth's motion through the ether, was
troubling the founders of modern physics\cite{1}. The puzzle stemmed, not from
astrophysical observations, but from the innocent question about the amount of
energy required to raise the temperature of diamond; that is, its specific
heat. The trouble was that much less energy was needed than expected. Almost a
hundred years before, Pierre-Louis Dulong and Alexis-Th\'er\`ese Petit
had observed 
that no matter what a solid is made of, its specific heat per molecule is
roughly the same. In 1876 this was put on an apparently solid theoretical
foundation by Ludwig Boltzmann. By applying his statistical mechanics to the
atoms in a solid he was able to compute the specific heat per atom in precise
agreement with Dulong and Petit's early observations. But by the last years of
the nineteenth century this triumph of statistical physics was looking
increasingly hollow. Not only did some materials, such as diamond, show too
small a specific heat but the advent of cryogenic techniques showed specific
heats to be strongly temperature dependent at low temperatures, in contrast to
Boltzmann's theory. Not for the last time would low-temperature physics point
to the need for a new understanding.

Attempts to resolve the clash between
theory and experiment were espoused by three great physicists of the
nineteenth century: Boltzmann, Lord Kelvin and Lord Rayleigh. Boltzmann
suggested that the way atoms behaved when confined within a solid might not be
as straightforward as he had assumed. Lord Kelvin, on the other hand, looked
to the mathematics of Boltzmann's derivation, convinced that the error lay
there. It was Lord Rayleigh who was prepared to say that both the experiment
and the theory were correct and that there was a true crisis which could only
be resolved by fresh insight. The next step to solving the problem was made by
Einstein.

\section{Einstein and the puzzle of specific heats}

Einstein took his
inspiration from an unlikely source---the light from stars. In 1900
Max Planck had developed a formula that could relate the colour of
stars to their temperature---the first step in what would become
quantum theory. Planck's theory had led Einstein to conclude that
light is made up of discrete quanta or `photons'. In 1906 Einstein
applied Planck's formula to the vibrations inside matter which, he
reasoned, must also be made up of quanta---tiny wave packets of sound
that we now call `phonons'. At high temperatures, Einstein's theory
reverted to that of Boltzmann, but once the temperature dropped below
that required to excite phonons, the heat content dropped
dramatically. In his paper of November 1906 Einstein published one of
his few fits to data---comparing his theory to the specific heat 
of diamond---to demonstrate his solution of
a seventy year old puzzle. A new discipline of the quantum theory of solids
was born.

Einstein's 1906 theory
was necessarily incomplete but it contained the seeds of an impending
revolution. A first-principles derivation of Einstein's simple theory had to
wait until 1924 when it became the first problem to be solved by Werner
Heisenberg using the new `quantum mechanics'. By 1930 the approach pioneered
by Einstein had essentially wrapped up the specific heat capacity problem.

\section{Quantum criticality and the return of the puzzle.}

A century later, the very
same measurement of the specific heat of solids points to a new clash with our
theories of matter. The `bad actors' of the twenty-first century are man-made
crystals at the forefront of modern materials science. Physics has
traditionally focused on stable phases of matter, such as those shown by
superconductors, magnets or ferroelectrics, but with modern materials it is
possible to study unstable quantum phases of matter. Puzzling new behaviour
develops around the precarious point of instability between two stable phases
of matter: the `quantum critical point'. The unique properties that develop in
quantum critical matter have become a major focus of research in the past ten
years. Moreover, the discrepancies that are unfolding between established
theory and experiment leave us in a state of affairs that is curiously similar
to that which existed with diamond a hundred years ago.

As the materials of
our lives become more sophisticated---from complex plastics to advanced metal
ceramics---one might think that understanding their properties would require
knowing, in ever greater detail, the intricate complexities of electron and
atom motion. Alas, such a task is impossible; even the most advanced computers
are unable to cope with more than a few dozen electrons at once. Fortunately
however, as Boltzmann foresaw when considering diamond, collections of
particles can behave differently from isolated ones, and although their
individual motions are complex, their collective properties acquire
qualitatively new forms of simplicity. The appearance of magnetism, the
development of zero-viscosity states in so-called superfluids, and the
emergence of zero resistivity in a superconducting metal, are all examples of
new, yet simple, behaviour that arise from the collective motion of electrons
or atoms in complex matter. These are all examples of stable phases of
matter. Collective behaviour also becomes important at the unstable interface
between two stable phases of matter: the point of transformation is called a
`phase transition'.

Phase transitions are ubiquitous: from the crystallization of water into
snowflakes, the alignment of electron spins inside a ferromagnet, the
emergence of superconductivity in a cooled metal to the very formation of
space-time in the early Universe; all involve phase transitions. Despite their
diversity, different phase transitions often share many fundamental
characteristics. The specific heat when water turns to steam at a critical
pressure has exactly the same power-law dependence on temperature as that of
iron when it is demagnetized by having its temperature raised. Understanding
this universal behaviour, known as `critical phenomena', was a triumph of
twentieth century physics\cite{2}. One of the key discoveries was that the imminent
arrival of order at such continuous phase transitions is signalled by the
formation of short-lived droplets of nascent order that grow as the system is
tuned to the critical point. At the critical point, the material is spanned by
droplets of all sizes.

The melting of ice is, like most phase transitions,
caused by the increase in random thermal motion of the molecules which occurs
as the temperature is raised. The ordered arrangement of atoms that exists in
the solid cannot be sustained beyond a certain temperature and the crystal
melts. Yet research into condensed matter over the past decade has revealed a
new kind of phase transition that is driven, not by thermal motion, but by the
quantum fluctuations associated with Heisenberg's uncertainty principle. These
quantum fluctuations are called `zero-point motion'. According to the
uncertainty principle, the more certain a particle's position, the more
uncertain is its velocity. Thus, even when random, thermal motion ceases at
the absolute zero temperature, atoms and molecules cannot be at rest because
this would simultaneously fix their position and velocity. Instead they adopt
a state of constant agitation. Like thermal motion, if zero-point motion
becomes too wild, it can melt order, but in this case the melting takes place
at absolute zero. Such a quantum phase transition\cite{3} takes place in solid
helium, which is so fragile that it requires a pressure to stabilize its
crystal lattice even at absolute zero. When the pressure is released,
zero-point motions melt the crystal.

The best studied examples of quantum phase transitions involve
magnetism in metals. Electrons have a magnetic direction or spin,
which when aligned in a regular fashion makes a material
magnetic. Iron magnetizes when all the spins inside align in parallel,
but in other materials the spins form a staggered, alternating, or
antiferromagnetic, arrangement (Fig. 1). These more fragile types of
order are more susceptible to melting by zero-point
fluctuations. Almost three decades ago theoretical physicist John
Hertz, now at Nordita, made the first study of how quantum mechanics
would affect phase transitions\cite{4}. Hertz was fascinated by the
question of how critical phenomena might be altered by quantum
mechanics. Applying quantum mechanics to phase transitions turns out
to be very like Einstein's relativistic unification of space-time. In
Hertz's theory, quantum mechanics appears by including a time
dimension to the droplets of nascent order. Normally this produces no
additional effect, but Hertz reasoned that if a phase transition took
place at absolute zero, then the droplets of order that foreshadow the
transition would become quantum-mechanical rather than classical. At a
zero-temperature phase transition, he reasoned, these quantum droplets
would grow to dominate the entire material, changing its properties in
measurable ways---and most affected would be the electrons
(Fig. 2). Such `quantum critical matter' offers the real prospect of
new classes of universal electronic behaviour developing independently
of the detailed material behaviour, once the material is driven close
to a quantum critical point\cite{5}.

\fg{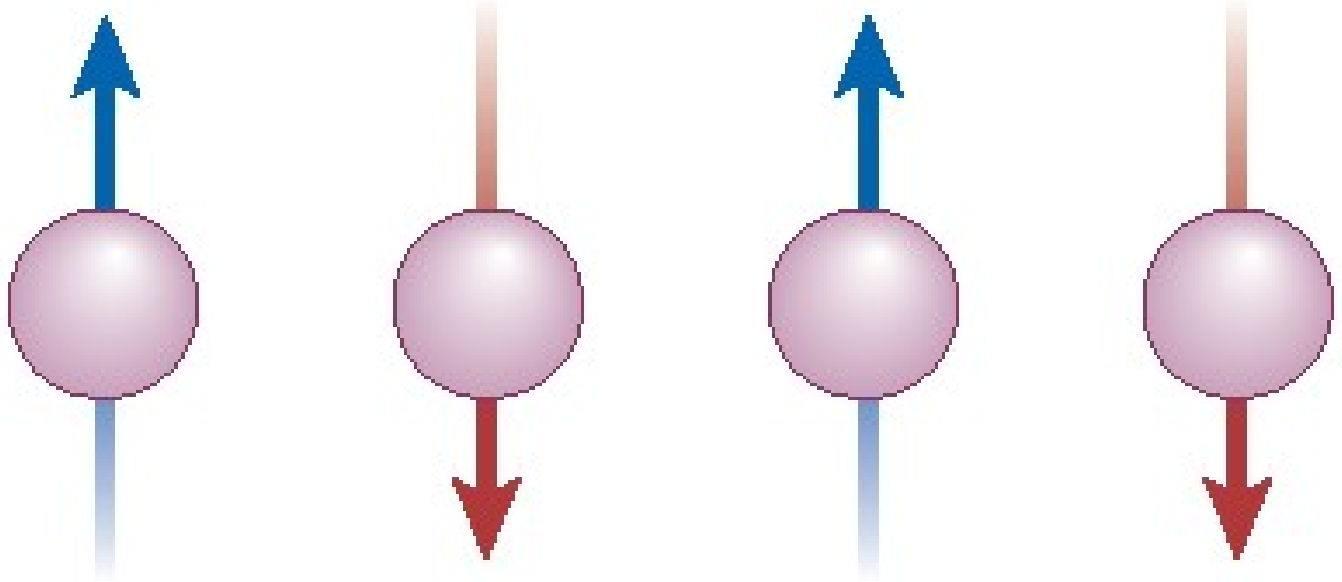}{fig1}{Staggered arrangement of spins in an antiferromagnet.}

The electrons that carry current in a metal are similar to photons of light:
they are quantum waves whose wavelength decreases as their momentum
increases. Unlike photons however, electrons obey the `Pauli exclusion
principle': no two electrons can share the same wavelength (or momentum). To
minimize their energy, each electron must occupy the momentum state with
lowest energy that is not already filled. When all electrons have done this,
there is then a sharp demarcation (called the Fermi surface) separating the
highest energy occupied and lowest-energy unoccupied momentum states. This
ordering imposes severe constraints: like apples packed into a barrel, only
those near the top can be easily rearranged. One consequence is seen in the
specific heat: only electrons near the Fermi surface can absorb thermal energy
and find an unoccupied momentum state to move to. This means that the specific
heat of metals is tiny, but that it grows linearly with temperature. The
coefficient of linearity, called the Sommerfeld coefficient, is a rough
measure of the `effective' mass of the highest-momentum electrons inside the
metal. When the effective mass is compared to that of an isolated electron it
is usually found to be somewhat larger. This is because of the forces between
electrons, which mean that as an electron moves around it has to push
neighbouring electrons aside.

\figwidth=9cm

\fg{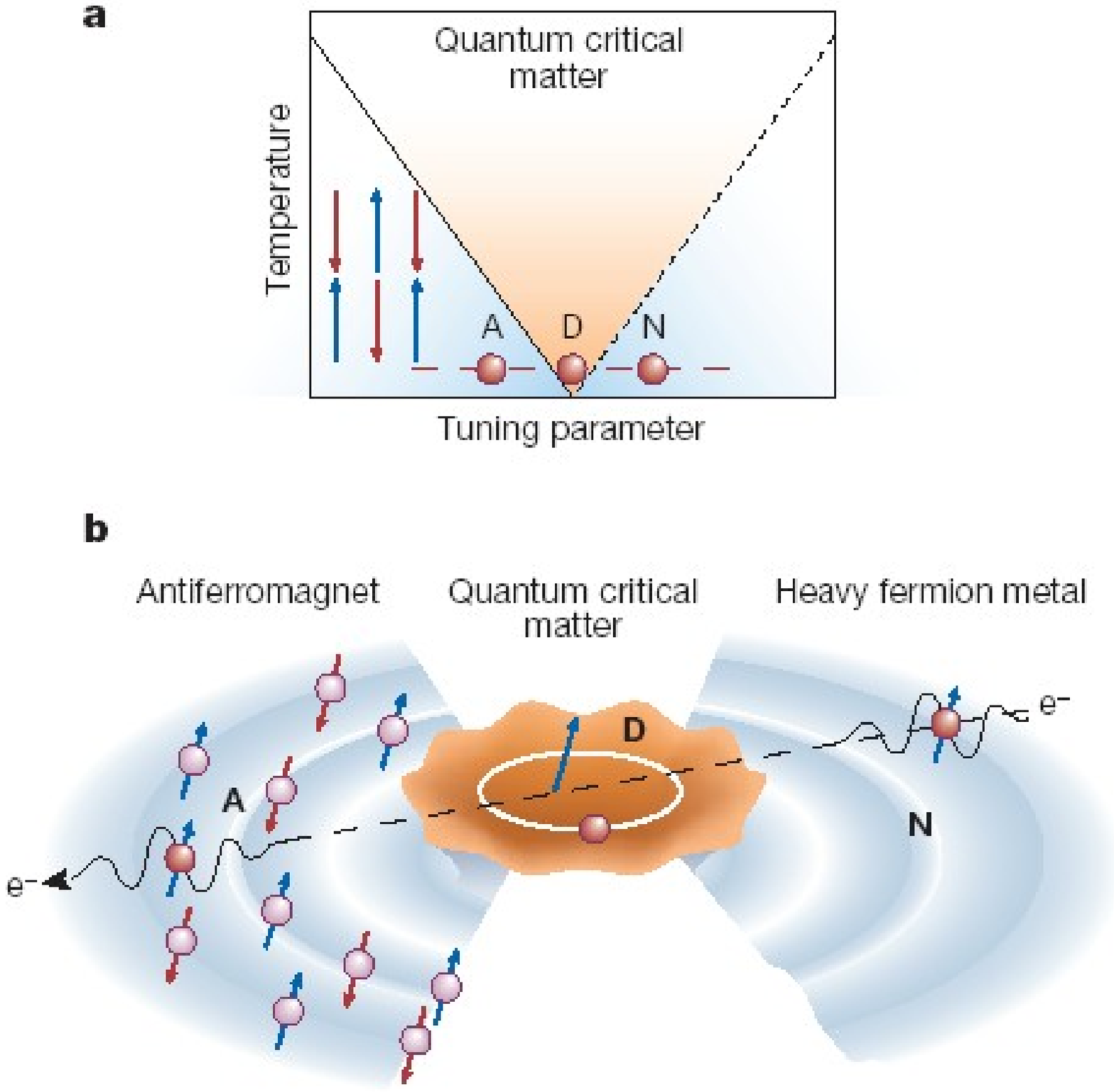}{fig2}{Schematic illustration of a quantum
critical point showing the phase diagram (a) and the growth of droplets of
quantum critical matter near the quantum critical point (b). a, Schematic
phase diagram near a quantum critical point. Quantum critical points distort
the fabric of the phase diagram creating a `V-shaped' phase of quantum
critical matter fanning out to finite temperatures from the quantum critical
point. b, As matter is tuned to quantum criticality, ever-larger droplets of
nascent order develop. On length-scales greater than these droplets, electrons
propagate as waves. Inside the droplet, the intense fluctuations radically
modify the motion of the electron, and may lead to it breaking up into its
constituent spin and charge components. Physics inside the V-shaped region of
the phase diagram (a) probes the interior of the quantum critical points (D),
whereas the physics in the normal metal (N) or antiferromagnet (A) reflects
their exterior. If, as we suspect, quantum critical matter is universal, then
no information about the microscopic nature of the material penetrates into
the droplets. Making an analogy with a black hole, the passage from
non-critical, to critical quantum matter involves crossing a `material event
horizon'. Experiments that tune a material from the normal metal past a
quantum critical point force electrons through the `horizon' in the phase
diagram, into the interior of the quantum critical matter, from which they
ultimately re-emerge through a second horizon on the other side into a new
universe of magnetically ordered matter.}

But much more dramatic changes occur near a quantum phase transition. Ten
years ago, Hilbert von Lohneyson's group in Karlsruhe, Germany, decided to
measure the specific heat of a quantum critical metal\cite{6}. They chose $\rm CeCu_{6}$,
which can be tuned to a magnetic---actually, an antiferromagnetic---quantum
critical point by adding small amounts of gold. As they added gold to the
metal, they found that the Sommerfeld ratio of the metal kept getting bigger
-- as if the approach to criticality led to a metal where the electrons were
getting heavier and heavier. But at the quantum critical concentration, the
Sommerfeld coefficient never settled down to a constant. It just kept on
rising as they lowered the temperature, as if the mass of the electrons on the
Fermi surface was becoming infinite and the energy of the electrons
vanishing. The Karlsruhe group found another disturbing property. In normal
metals, the electrical resistance due to electrons scattering off one another
grows as the square of the temperature, but in this system, at the quantum
critical point, the resistivity is linear with temperature. The constancy of
the Sommerfeld coefficient and the quadratic temperature dependence of the
resistivity constitute two absolutely stalwart signatures of normal
metals. Their breakdown suggests that a quantum critical metal is a
fundamentally new type  of electron fluid. Since the original Karlsruhe
measurements, many new discrepancies have come to light\cite{7}. Quantum critical
points have now been found by pressure tuning\cite{8} and by applying magnetic
fields\cite{9}. We even have an apparent case of a line of quantum criticality
instead of a single point\cite{10}. All support the notion that the characteristic
energy scale of the metal is driven to zero at a quantum critical
point. Indeed, temperature itself seems to be the only energy scale that
remains in quantum critical matter. In neutron scattering\cite{11}, for example, the
rate at which electrons are scattered off the critical magnetic fluctuations
seems to depend solely on the ratio of energy to temperature. Linear, or
quasi-linear, resistivity is another sign of this phenomenon and, in the more
dramatic cases, it can be followed over three decades in temperature. 

\figwidth=10cm

\fg{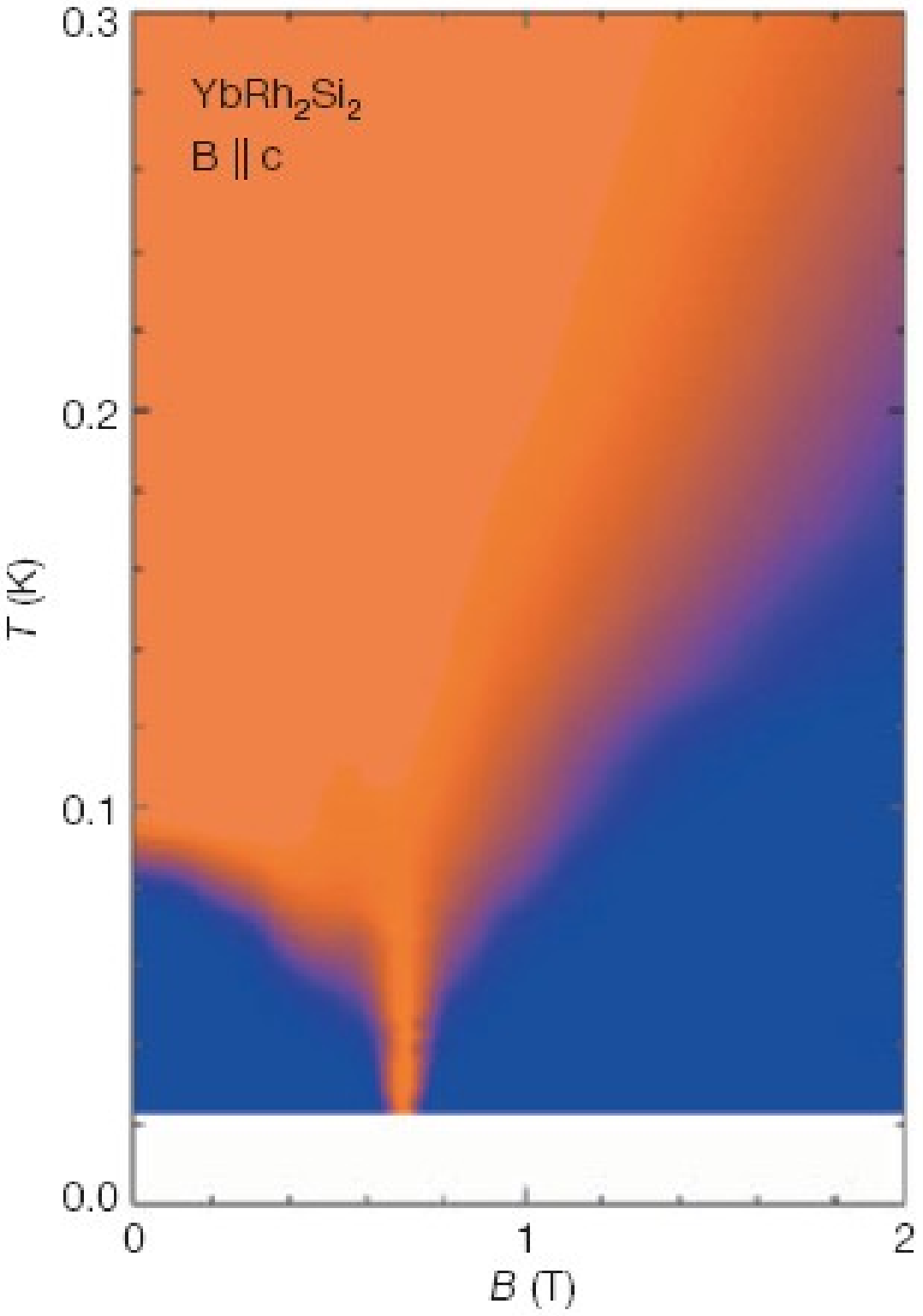}{fig3}{`Singularity in the phase diagram' illustrated by
data taken from the material $\rm YbRh_{2}Si_{2}$ where an applied magnetic field tunes
the material to a quantum critical point[25]. Blue regions indicate normal
metallic behaviour. Orange regions indicate anomalous metallic behaviour with
linear resistivity. The singular quantum critical point at absolute zero
produces a wide region of unusual metallic behaviour at finite temperatures.}

One
thing becoming increasingly clear is that although matter can never be cooled
down to the quantum critical point at absolute zero, 
drastic effects are felt long before this point is reached (Figs 2 and
3). It is this influence that elevates quantum criticality from an
intellectual abstraction at absolute zero to a real-world phenomenon
that can profoundly change finite-temperature material properties. The
quantum critical point represents a kind of `black hole' in the
material phase diagram and this proves a useful analogy. Just as the
black hole of cosmology distorts surrounding space-time, quantum
critical points distort the fabric of the phase diagram (Fig. 2),
creating a `V-shaped' region of quantum critical matter fanning out to
finite temperatures from the quantum critical point.

\section{Resolving the quantum critical enigma}

Some of the strange properties we see in quantum critical metals echo the
predictions of Hertz and subsequent extensions of his theory\cite{12}, but this
conventional wisdom is flouted in at least two important respects. First, the
effects are much stronger and wide ranging than Hertz expected. Second, the
particular case of antiferromagnetism should be by far the least dramatic: the
alternating pattern of up and down magnetism should average out for almost all
electrons, slowing down only the tiny fraction that can interfere
constructively with this arrangement. In marked contrast, the experiments tell
us that the effective mass of every electron on the Fermi surface is being
increased to infinity---in essence, bringing every electron to a halt. Like
Einstein a century before us, we have a thirty-year-old theory that cannot
account for present experiments.

The current theoretical alternatives also in
some sense parallel those of Einstein's contemporaries, and there is great
controversy. Some believe that the Hertz theory can indeed be saved if proper
account is taken of the complexity of the material. The Karlsruhe team have
advanced the idea that the underlying antiferromagnetic spin fluctuations at a
quantum critical point collectively arrange themselves to become
two-dimensional\cite{13}. In this case, the motion of electrons between magnetic
layers becomes highly turbulent, and it is possible to understand how they are
driven to a halt in quantum critical matter. The only problem is that there is
no obvious way for the spin fluctuations to become stubbornly two-dimensional
in metals like $\rm CeCu_{6}$ where the electrons are not confined to layers.

Others
seek to explain the discrepancy as a failure of Hertz's original
approximations\cite{14}. They point out that the distinction between the fluctuating
magnetism and the electrons that appears in the Hertz theory is not so obvious
when you remember that the magnetism itself also comes from the same
electrons. Add to that the realities that materials contain disorder (atoms
out of place, for example) and the assumptions of the mathematical derivations
may not be quite so convincing.

Finally, there is the third possibility that,
as Lord Raleigh advocated in the case of diamond, there is a true crisis and a
new framework for quantum phase transitions is required. That would certainly
be remarkable since the current theory is a beautiful synthesis of two very
successful foundation stones of modern physics: the theory of
temperature-driven phase transitions and quantum mechanics. Yet in the face of
such puzzling experimental observations we are being driven towards such a
possibility.

Several fascinating ideas have come to the forefront. One
suggestion is that we are seeing a new kind of quantum criticality that turns
classical criticality on its head. Classical criticality involves the global
growth of static order in space: Hertz's theory takes this global model and
also includes growth along the quantum time dimension. An alternative proposal
is that quantum critical matter involves the growth of droplets of order in
time, but not space\cite{15}. The debate between proponents of `global' and this
`local' quantum critical theory is highly contentious. Another possibility is
that electrons break up inside the highly collective environment of quantum
critical matter. It could be that electrons inside quantum critical matter
break up into their constituent spin and charge components\cite{16}---like atoms
splitting up into ions in solution. This is related to `deconfined
criticality' which has been used to describe quantum critical points in
magnetic, two-dimensional insulators\cite{17}. The only problem is that no one yet knows how to
apply these ideas in detail to truly three-dimensional quantum critical
metals.

The puzzle of quantum criticality is tantalizingly important for
material science because, quite frequently, electrons appear to preempt the
intense critical fluctuations of a quantum critical point by re-organizing
themselves at the last minute into a new stable phase of matter. The
divergence of the effective masses of the electrons and the collapse of
electron kinetic energies all point to the formation of a highly degenerate
state at the quantum critical point making them very susceptible to
transformation into alternative stable electronic configurations.

For this
reason, quantum criticality may be a highly effective catalyst for the
formation of new stable types of material behaviour, providing an important
new route for the design and discovery of new classes of material. One
tendency is towards superconductivity\cite{18,19}. High-temperature superconductors
-- which currently become superconducting at liquid nitrogen temperatures and
inspire the hope for one day achieving room-temperature superconductivity --
exhibit a linear resistivity like metals, up to their melting
temperature. Some believe that high-temperature superconductors and a whole
host of other superconductors with quasi-linear resistances in the normal
state, are driven by quantum criticality. Paradoxically, it is hard to prove
this hypothesis because any attempt to remove the surrounding order will, at
the same time, destroy the quantum criticality. We are faced by what has been
described as a `quantum conundrum' \cite{5}.

Many other unidentified phases also appear to develop around quantum
critical points. In $\rm Sr_{3}Ru_{2}O_{7}$(ref. \cite{20}) and
$\rm URu_{2}Si_{2}$(ref. \cite{21}), magnetic fields are used to tune the
materials to a quantum critical point. These magnetic fields are too
strong to allow the electrons to use superconductivity to avoid the
critical singularity. Yet, rather than face the quantum critical point
itself, the electrons find another way out. Abrupt changes in the
resistivity and even in the material shape signal that the electrons
have adopted a new type of ordered state. Is this a new form of
magnetic order\cite{22} or a re-arrangement of the electrons'
flow\cite{23} or perhaps both?  All we know is that the electrons are
still mobile but the precise rules of their new collective motion are
still to be fully characterized and understood.

A hundred years ago, Einstein looked to the stars for inspiration
to understand the properties of cold, stable, quantum matter. Today, the
mysterious properties of critically unstable quantum matter not only suggest a
new pathway to new material design, but they also raise our hopes for a new
link between matter in the laboratory and matter in the cosmos. It has
recently been proposed that ideas of quantum criticality may link up with
quantum gravity at the surface of a black hole\cite{24}. Moreover, the break-up of
electrons inside quantum critical matter and the emergence of new forces
closely parallels challenges faced by the particle physicist, who seeks to
understand how new forces develop between particles as they break up at high
energies. Just as the inspiration of photons from stars led Einstein to solve
the mystery of specific heat in stable quantum matter a hundred years ago,
some of the ideas now being considered for the properties of particles in the
early Universe, such as gauge theory and supersymmetry, may be poised to
reappear, right under our nose in the laboratory, as the solution to the
specific heat problem in quantum critical matter.

We gratefully acknowledge discussions with P. Chandra, Z. Fisk,
A. P. Mackenzie and D. Pines. P.C. is supported by the National
Science Foundation grant NSF DMR 0312495. A.J.S. is supported by the
Royal Society, the Leverhulme Trust and the EPSRC. 
\bigskip

\noindent Competing interests statement: The authors declare that they have no
competing financial interests.



\begin{thebibliography}{99}

\bibitem{1} Pais, A. {\it Subtle is the Lord: the Science and the Life of
Albert Einstein}, Ch. 20. 389-401 (Oxford Univ. Press, Oxford, 1982). 

\bibitem{2} Domb, C. {\it The Critical
Point: a Historical Introduction to the Modern Theory of Critical Phenomena}
(Taylor \& Francis, London, 1996). 

\bibitem{3} Sachdev, S. {\it Quantum Phase Transitions}
(Cambridge Univ. Press, New York, 1999) 

\bibitem{4} Hertz, J. Quantum critical
phenomena. {\it Phys. Rev. B} {\bf 14}, 1165--1184 (1976). 

\bibitem{5} Laughlin, R. B., Lonzarich,
G. G., Monthoux, P. \& Pines, D. The quantum criticality
conundrum. {\it Adv. Phys.} {\bf 50}, 361--365. (2001). 

\bibitem{6} von Lohneysen, H. {\it et
al.}, Non-Fermi-liquid behavior in a heavy-fermion alloy at a magnetic
instability. {\it Phys. Rev. Lett.} {\bf 72}, 3262--3265 (1994). 

\bibitem{7} Stewart,
G. R. Non-Fermi-liquid behavior in d- and f-electron
metals. {\it Rev. Mod. Phys.} {\bf 73}, 797--855 (2001). 

\bibitem{8} Julian, S. R. {\it et al.} The normal
states of magnetic $d$ and $f$ transition metals. {\it 
J. Phys. Condens. Matt. }
{\bf 8}, 9675--9688 (1996). 

\bibitem{9} Grigera, S. A. {\it et al.} 
Magnetic field tuned quantum criticality in the metallic ruthenate
$\rm Sr_{3}Ru_{2}O_{7}$. {\it Science} {\bf 294}, 329--332 (2001).

\bibitem{10} Doiron-Leyraud, N. {\it et al.} Fermi liquid breakdown in the
paramagnetic phase of a pure metal. {\it Nature} {\bf 425}, 595--599 (2003). 

\bibitem{11} Schr\"oder,
A. {\it et al.}, Onset of antiferromagnetism in heavy-fermion
metals. {\it Nature} {\bf 407}, 351--355 (2000).

\bibitem{12} Millis, A. J. Effect of a non-zero temperature on quantum
critical points in itinerant fermion systems. {\it Phys. Rev. B} {\bf
48}, 7183--7196 (1993).

\bibitem{13} Rosch, A. Interplay of disorder and spin fluctuations in the
resistivity near a quantum critical point. {\it Phys. Rev. Lett.} {\bf
82}, 4280--4283 (1999).

\bibitem{14} Belitz, D., Kirkpatrick, T. R. \& Roll\" uhler, J. 
Breakdown of the perturbative renormalization group at certain quantum
critical points. {\it Phys. Rev. Lett.} {\bf 93}, 155701/1--4 (2004).

\bibitem{15} Si, Q., Rabello, S.,
Ingersent K. \& Smith, J. L. Locally critical quantum phase transitions in
strongly correlated metals. {\it Nature} {\bf 413}, 804--808 (2001). 

\bibitem{16} Coleman, P.,
P\' epin, C., Qimiao Si \& Ramazashvili, R. How do Fermi liquids get heavy and
die? {\it J. Phys. Condens. Matt.} {\bf 13}, R723--R738 (2001). 

\bibitem{17} Senthil, T.,
Vishwanath, A., Balents, L., Sachdev, S. \& Fisher, M. P. A. Deconfined quantum
critical points. {\it Science} {\bf 303}, 1490--1494 (2004). 

\bibitem{18} Mathur, N. D. {\it et
al.} Magnetically mediated superconductivity in heavy fermion
compounds. {\it Nature} {\bf 394}, 39--43 (1998).

\bibitem{19} Petrovic, C. {\it et al.} A new heavy-fermion superconductor
$\rm CeIrIn_{5}$: a relative of the cuprates? {\it Europhys. Lett.} {\bf
53}, 354--359 (2001).

\bibitem{20} Grigera, S. A. {\it et al.} Disorder-sensitive phase formation linked to
metamagnetic quantum criticality. {\it Science} {\bf 306}, 1154--1157 (2004). 

\bibitem{21} Kim, K. H., Harrison, N., Jaime, M., Boebinger, G. S. \& Mydosh,
J. A. Magnetic-field-induced quantum critical point and competing
order parameters in $\rm URu_{2}Si_{2}$. {\it Phys. Rev. Lett.} {\bf
91}, 256401/1--4 (2003).

\bibitem{22} Amitsuka, H. {\it et al.} Hidden order and weak
antiferromagnetism in $\rm URu_{2}Si_{2}$. {\it Physica B}
{\bf 312--3}, 390--396 (2002). 

\bibitem{23} Chandra, P. {\it et al.} Hidden orbital order in
$\rm URu_{2}Si_{2}$. {\it Nature} {\bf 417}, 831--834 (2002). 

\bibitem{24} Chapline, G. \& Laughlin, R. B. in
{\it Artificial Black Holes} (eds Novello, M. {\it et al.}) 179--198 (World
Scientific, Singapore, 2002). 

\bibitem{25} Custers, J. {\it et al.} The break up of heavy electrons at a
quantum critical point. {\it Nature} {\bf 424}, 524--527 (2003).

\end{thebibliography}
\end{document}